# A Multi-View Learning Approach to Enhance Automatic 12-Lead ECG Diagnosis Performance

Jae-Won Choi[†]
Infomining, Co. Ltd.
Gyeonggi, Republic of Korea
moam1127@infomining.co.kr

Dae-Yong Hong[†]
Infomining, Co. Ltd.
Seoul, Republic of Korea
dyhong@infomining.co.kr

Chan Jung[†]
Infomining, Co. Ltd.
Incheon, Republic of Korea
chan@infomining.co.kr

Eugene Hwang
KAIST College of business
Seoul, Republic of Korea
hegene3686@kaist.ac.kr

Sung-Hyuk Park
KAIST College of business
Seoul, Republic of Korea
sunghyuk.park@kaist.ac.kr

Seung-Young Roh
Korea University College of Medicine and Korea University Medicine
Seoul, Republic of Korea
syroh@korea.ac.kr

## ABSTRACT

The performances of commonly used electrocardiogram (ECG) diagnosis models have recently improved with the introduction of deep learning (DL). However, the impact of various combinations of multiple DL components and/or the role of data augmentation techniques on the diagnosis have not been sufficiently investigated. This study proposes an ensemble-based multi-view learning approach with an ECG augmentation technique to achieve a higher performance than traditional automatic 12-lead ECG diagnosis methods. The data analysis results show that the proposed model reports an F1 score of 0.840, which outperforms existing state-of-the-art methods in the literature.

## 1 INTRODUCTION

Cardiovascular diseases account for one-third of annual deaths worldwide [1]. An electrocardiogram (ECG), which records the electrical activity of the heart as a waveform, is the most basic test available to check the heart health, which in turn is critical for the management and early diagnosis of cardiovascular diseases. However, a high level of expertise is required for accurate reading of an ECG. Furthermore, the large number of ECG records, over 300 million cases worldwide every year, places a huge burden on medical staff [2]. To address this issue, the 12-lead ECG device provides algorithm-based computerized interpretation of ECG, which shortens the reading time by 24-28% [3]. However, this suffers from a limitation of a high false-positive rate. This has given rise to the need for fast and accurate computer-aided automated ECG interpretation.

There are three main streams of research on automatic ECG diagnosis. The first is handcrafted feature–based conventional machine learning research [4–7]. However, owing to ECG artifacts and individual differences, these are difficult to generalize and cannot be configured in an end-to-end manner. The second is mathematical modeling–based research. Attempts have been made to model cardiac dynamics by analyzing ordinary differential equations. However, strict formalization is challenging owing to the numerous variables and inherent nonlinearity of ECG [8].

Finally, deep learning (DL)–based studies have recently attracted considerable attention. DL has achieved remarkable success in general tasks over the past decade [9] and its potential as a diagnostic aid in the medical field is increasing [10–12]. DL components, such as convolutional neural networks (CNNs) [13–17], long short-term memory (LSTM) [18–21], and attention [22], that excel in general tasks in within the medical field have been adopted. Considering that ECG, like bio-signals, periodic signals, time series, and multi-sensory data, is a complex signal, the integration of more varied techniques has been explored [23]. As such, the modern trend of DL-based ECG diagnosis research is to understand ECG more deeply by combining various DL components [24–28]. Table 1 presents a summary of DL-based studies. To the best of our knowledge, the study by Chen et al. achieved the best F1 score among comparable studies on the CPSC2018 database [32].

| Researchers | Preprocessing technique | Modeling technique | F1 score |
|---|---|---|---|
| He et al. [23] | Oversample, zero-pad, signal segmentation | CNN, Bi-LSTM | 0.806 |
| Luo et al. [24] | Median filter, signal segmentation | CNN, Bi-LSTM | 0.822 |
| Ye and Lu [27] | Oversample, DB6 wavelet, signal segmentation | CNN, Bi-LSTM, XGBoost | 0.812 |
| Wang et al. [19] | Augmentation, replicate, cutout to 10s | Deep Multi-Scale Fusion CNN | 0.828 |
| Ge et al. [28] | Zero-pad, filtering (Butterworth vs wavelet transform) | CNN, SE block | 0.828 |
| Chen et al. [26] | Zero-pad | CNN, Bi-RNN, Attention | **0.837** |
| Zhang et al. [20] | Augmentation (random scaling and shifting) | CNN | 0.813 |
| Ye et al. [31] | Augmentation, feature fusion (DNN and patient information) | CNN, LSTM, XGBoost | 0.818 |

**Table 1. Summary of DL based methods.** Preprocessing techniques and DL modules used in previous works. F1 scores shown are from each literature. Not all performances are compatible because their datasets differ. Luo et al., Chen et al., and Zhang et al. utilized the CPSC2018 database on which our experiments are conducted. Descriptions for the database is in Section 2.1 and 5.1. The best F1 score is written in bold.

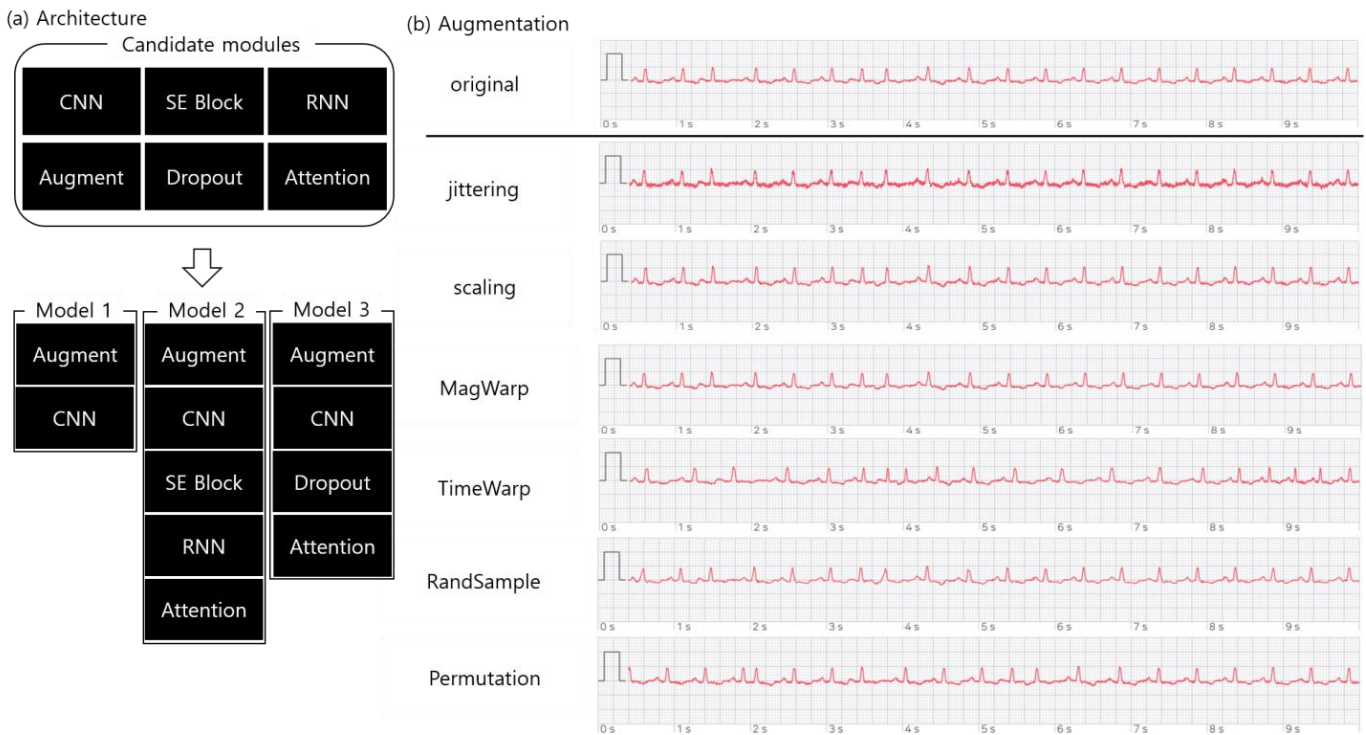

**Figure 1. Overall architecture and augmentation for ECG.** (a) Combination of components for various models. (b) ECG-specific augmentation

We focused on the combination of various components in previous DL-based studies and found that most explored a single way of combining specific components. Therefore, we investigated multiple methods of integration of more diverse DL components based on multi-view learning [29]. Candidate components with outstanding practical performance are listed, and combinations are compared to investigate their better performance (Figure 1a).

There are six types of artifacts in the ECG [30], but many studies have not focused on them [13–15, 18, 26, 27]. In this study, we propose an ECG-specific data augmentation technique to simulate various ECG artifacts (Figure 1b). In addition, the degree and number of transformations are fully random. This provides diverse data to the model so that more general representations can be learned.

During training, every single model was fed 30 s of 12-lead ECGs as input. After convergence, they become a candidate set for constructing ensemble models. For inference, a varied-length ECG signal can be inferred by averaging. We achieved state-of-the-art performance (average F1 score of 84.00%) on CPSC2018 by using an ensemble model.

The characteristics of ECG and the relationship between each DL component and ECG are described in Section 2, and ECG augmentation is described in Section 3. Section 4 explains the working of the combination of techniques and models. Sections 5 and 6 introduce the experiments and the analysis of the results, respectively.

The main contributions of this study are as follows:
(1) We propose a data augmentation technique specialized for ECG and induce models to perform robust ECG interpretation.
(2) With respect to multi-view learning, we investigate various architectures through a combination of DL components, and confirm that by synthesizing them, improved performance can be achieved.

## 2  DL Components and Their Roles in ECG Diagnosis

### 2.1  ECG Datasets

Prior to the 1990s, researchers constructed databases such as MIT-BIH [33], PTB [34], and INCART [35] for automatic ECG diagnosis. However, most were small-scale and contained fewer than 12 leads. With the recent advent of the DL approach, large-scale databases such as CPSC2018 and PTB-XL [36] have emerged. We experimented with the CPSC2018 dataset using varied-length ECGs with more abundant target diagnoses. More detailed information on CPSC2018 is presented in Section 5.1.

### 2.2  Characteristics of ECG

In automatic ECG diagnosis, multiple factors should be considered. Morphological information is extracted from the shape of the waveform (P, QRS, and T). Temporal information is obtained by referencing the intervals between the waveforms or peaks. The dependence between points in each lead or between leads must be considered because the lead in which a certain pattern appears specifies the heart regions in which a lesion has occurred. Furthermore, in their study, Liu et al. [23] highlighted three more characteristics of ECG: integrity, diversity, and periodicity. Therefore, various types of information should be considered in the automatic ECG diagnosis.

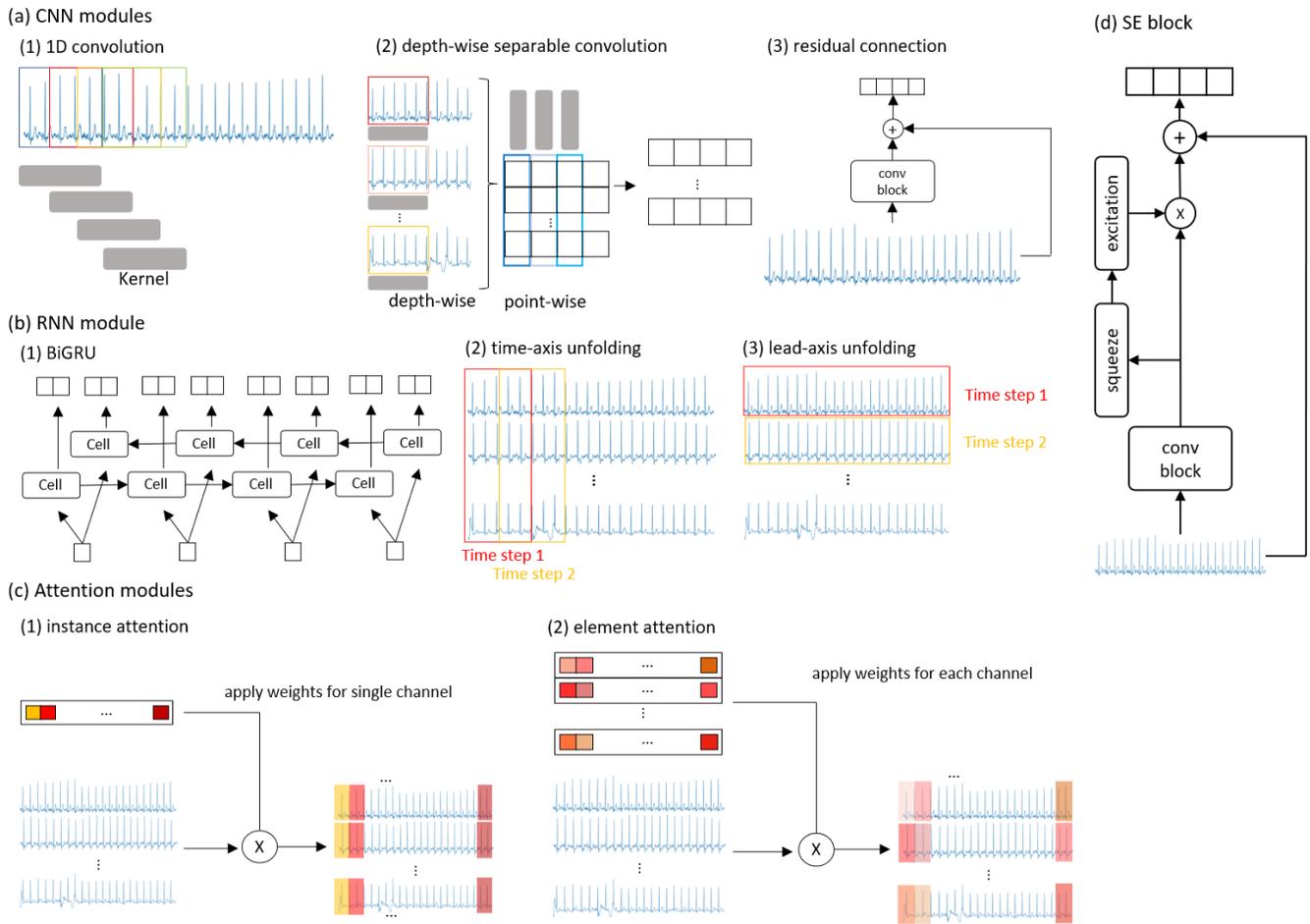

**Figure 2. The detailed structure of each component used in the experiments.** (a) CNN modules (b) RNN modules (c) Attention modules (d) SE block

### 2.3　CNN

CNNs efficiently extract spatial features and are parameter-efficient. In this study, a CNN was adopted as the basic component (Figure 2a). 1D convolution (Figure 2a1) and depth-wise separable convolution [37] were used to separate the features for each channel, increasing the computational efficiency (Figure 2a2). Referring to the effect of the residual connection in ECG [20, 38], ResNet was chosen as the default architecture (Figure 2a3).

### 2.4　RNN

A recurrent neural network (RNN) is effective for reflecting the sequential characteristics of a time series. The experiment in this study utilized a gated recurrent unit (GRU) [39] for computational efficiency. The GRU operates as follows:

$$r_t = sigmoid(W_r x_t + U_r h_{t-1}) \tag{1}$$

$$h_t = z_t h_{t-1} + (1 - z_t)\tilde{h}_t \tag{2}$$

$$\text{where } z_t = sigmoid(W_z x_t + U_z h_{t-1})$$

$$\tilde{h}_t = \tanh(W x_t + U(r_t * h_{t-1}))$$

$W_r, W_z, W, U_r, U_z$, and $U$ are trainable parameters. Formula 1 represents a reset gate, and Formula 2 represents an update gate. Considering that ECG does not have unidirectional dependency, a bidirectional gated recurrent unit (BiGRU) was used as a component (Figure 2b1). The 12-lead ECG consists of a temporal axis (Figure 2b2) and a spatial axis (Figure 2b3). Therefore, we considered the cases of time-axis unfolding and lead-axis unfolding.

### 2.5　Attention

Attention introduces weights to the encoding process in the neural machine translation (NMT) task [40], emphasizing key information. The experiment was performed as shown in Figure 2c1, according to the method in [26]. In addition, to consider the inter-lead difference, the point-wise attention (element attention) method was added (Figure 2c2). The operation of the attention layer is as follows:

$$u = \tanh(W_1 X^T), \quad X \in \mathbb{R}^{L \times C}, W_1 \in \mathbb{R}^{C \times C} \tag{3}$$

$$a = softmax(W_2 u), \quad W_2 \in \mathbb{R}^{1 \times C} \text{ or } \mathbb{R}^{C \times C} \tag{4}$$

$$z = \sum_{i=1}^{L} a_i^T \cdot x_i \tag{5}$$

For simplicity, the bias was omitted. $X$ is the input, $W_1$ and $W_2$ are trainable parameters, $a$ is the weight to be applied, $z$ is the output, $L$ is the input length, and $C$ is the number of channels. $W_2 \in \mathbb{R}^{1 \times C}$ for instance attention and $W_2 \in \mathbb{R}^{C \times C}$ for element attention.

### 2.6　SE Block

The squeeze and excitation network (SENet) [41] rescales the channel-wise feature response based on its importance to the task. The SE operations, which summarize the feature map and importance, respectively, can achieve a high-performance improvement with a small parameter increase (Figure 2d). Ge et al. [25] introduced an SE block for ECGs and confirmed that the degree to which each character is reflected in the target diagnosis can be well readjusted.

## 3　Data Augmentation for ECG

Because data augmentation must preserve labels, it is difficult to apply computer vision techniques, such as vertical flipping. Therefore, referring to seven augmentation techniques [42] for Parkinson's disease data, we propose an ECG augmentation technique (Figure 1b). Parkinson's disease data is a 3-channel biosignal, and it is closely related to ECG in that it measures the electrical signal of the brain while ECG measures that of the heart.

The proposed random augmentation method is based on the following six operations.

- Jittering is applied by additive high-frequency noise and represents power line interference (PLI).
- Scaling introduces electromyography (EMG) noise by adjusting the amplitude of the signal.
- Magnitude warping simulates baseline wander by distorting the magnitude of an arbitrary location.
- Time warping provides small changes to each point of the ECG along the x-axis; therefore, the interval characteristics of signal components such as PR, RR, or QT intervals.
- Permutation prevents position memorization by reversing positions between arbitrary sections within the ECG.
- RandomSample re-samples based on non-uniform random points. The sampling rate was lowered by up to 20% from the original, representing different devices or environments. This may lead to a signal-quality degradation by losing a small number of signal details.

The parameters of each transformation were experimentally set according to the ECG characteristics. Parameters are not specified as constants but are randomly assigned for each mini-batch within a heuristically defined parameter space. The number of transformations to be applied is also randomly determined.

## 4　Combination of Components based on Multi-View Learning

Studies up to the early 2010s on DL-based techniques were interested in differences according to the model architectures or their hyperparameters. However, recent research focused on what is good representations and how to obtain them [43]. These representation-based techniques commonly focus on synthesizing features from various perspectives. One of these concepts is multi-view learning.

Multi-view learning trains multiple models to select different representations of data. Multi-view learning involves two basic principles [44]. The first is consistency, which states that the context should be consistent even if the views are different. In our experiment, each model had the same inference goal for the same task. The second is the complementary principle, which states that each view should be able to complement the other. The CNN feature considers the shape of a waveform while the RNN feature considers the temporal context. Therefore, the latter complements the former with temporal information.

Multi-view learning can be broadly classified into three types, data-level, classifier-level, and representation-level integration, according to when to integrate [45]. The data-level learning integrates different types of data in the embedding process and uses techniques such as principal component analysis (PCA). The classifier-level learning integrates the predictions of the models trained using different views. Ensemble is a representative example of a classifier-level scheme. The representation-level learning incorporates latent vectors and is often used in multimodal tasks. In this study, to utilize ECG augmentation and the representation ability of the model, an ensemble was constructed based on the second category. However, it is also related to the third category in that it induces fusion between representations from diverse leads through interactions between components. This layer configuration facilitates information exchange between the signal segments, enabling higher-level inference.

Because we focused on the difference in viewpoints between different components, models of various structures were assembled on the same data split. To fuse the model outputs, we averaged all probability vectors with the intention of equal contribution of all models. Our model obtained an accurate classification ability, noise immunity, and the ability for rich representation by synthesizing different central features.

## 5 Experiments

### 5.1 Dataset

CPSC2018 was used for the experiments. It consists of 6877 public ECG records and 2954 hidden ECG records. The database consists of 12-lead ECGs with a sampling rate of 500 Hz and includes eight normal arrhythmia classes (AF, I-AVB, LBBB, RBBB, PAC, PVC, STD, and STE). The signal lengths of the public records range from 6 s to 60 s. For the experiment, we used a data split of ratio 8:1:1, namely "*our training set*", "*our validation set*", and "*our test set*," to make it clear they are from the public records. The "*hidden test set*" in this literature indicates the ECG records that are not disclosed by the challenge organizer. Every model was trained on our training set, whereas the hyperparameters were tuned based on our validation set. Our test set was used for internal evaluation, and the final top 5 ensembles were used by the challenge organizer to evaluate the hidden test set for external validation. During the data split, all ECGs, including multiple labels, were included in the test set. The public dataset contains 476 multilabel records. For multi-label cases, the challenge's official evaluation defines it as sufficiently correct if the model's multiclass inference is included as one of its labels. Inappropriate data containing NaN were removed. The remaining were segmented into 30 s. If shorter, it was unified to 30s through zero-padding. No preprocessing but z-score normalization was applied.

| Condition | Cases |
|---|---|
| augment | Randomly applied (0 ~ 6 of 6 methods) |
| backbone | ResNet-18 / ResNet-34 (both with separable conv) |
| SE block | None / r=2 / r=4 / r=8 (r : reduction ratio) |
| Activation | ReLU / ELU / Leaky ReLU |
| BiGRU | None / time-axis unfolding / lead-axis unfolding |
| Attention | None / instance attention / element attention |
| Total | 216 structures |

**Table 2. Architecture Cases.** In **"**ResNet-XX**"**, XX indicates the number of convolution layers.

### 5.2 Settings

Experiments were performed using TensorFlow 2.5, on an Intel Xeon Gold 5218, GEFORCE 3090 24GB X 4, 192GB RAM, CentOS. The overall composition of the experiment is as follows: First, the architecture was constructed using a combination of various components (Table 2). Because all possible scenarios could not be tested, the combinations and parameter ranges were heuristically limited. Augmentation was always applied, but which operation to use, the number of operations, and parameters within each one were fully random in mini-batch units. The architectures were constructed according to five other conditions. For each architecture, trainable parameters were tuned based on the validation set in the search space. Based on our validation set, ten models showing optimal performance were adopted as members of the candidate set. An ensemble model was constructed based on a random number of models in the candidate set and the top five ensembles were selected. Finally, the performance of the CPSC2018 hidden test set for the best ensemble was evaluated. The details of the architecture case for calculating the top 10 single models are as follows.

The architecture follows the ResNet form. Because the ECG has a lower dimension than the images, the number of channels in all convolutional blocks is reduced by half. For the first layer, 16 kernels (reduced by half again), not 32 channels, were used, considering that the ECG has 12 channels. Each kernel size was three, and ResNet-18 and ResNet-34 were compared with the backbone. The reduction ratio r of the SE block was set to 2, 4, and 8, considering the trade-off between the complexity and performance. Activation compared ReLU, ELU, and leaky ReLU. BiGRU was added as a single layer with a dropout rate of 0.2, tanh activation, and 256-length input and output when used. Lead-axis unfolding (Figure 2b3) was implemented by transposing the input. The attention component was applied on top of the BiGRU layer; furthermore, instance or element attention was used. Because both include weighted summation (Formula 5), the GAP is added only when attention is not used. A softmax dense classifier follows at the end of the model.

| Model | Backbone | Augment | SE block | BiGRU | Attention | Activation | F1 |
|---|---|---|---|---|---|---|---|
| M01 | ResNet-18 | O | X | Lead | Element | ELU | **0.8456** |
| M02 | ResNet-18 | O | X | Lead | Element | ReLU | 0.8406 |
| M03 | ResNet-18 | O | X | Lead | Instance | ReLU | 0.8396 |
| M04 | ResNet-18 | O | X | Lead | Instance | ReLU | 0.8331 |
| M05 | ResNet-18 | O | r=2 | Lead | Element | Leaky ReLU | 0.8321 |
| M06 | ResNet-18 | O | r=2 | Lead | Instance | Leaky ReLU | 0.8296 |
| M07 | ResNet-18 | O | r=2 | Lead | Element | ReLU | 0.8259 |
| M08 | ResNet-18 | O | X | Lead | Element | Leaky ReLU | 0.8243 |
| M09 | ResNet-18 | O | r=2 | Lead | Instance | ELU | 0.8186 |
| M10 | ResNet-18 | O | r=2 | Lead | Element | ELU | 0.7970 |

**Table 3. Structure of Top10 Single Models.** "O" and "X" refer to "applied" and "not applied" respectively. "r" indicates the reduction ratio for the SE block. "Lead" stands for lead-axis unfolding described in Section 2.4. "Element" and "Instance" refer to inter-lead and intra-lead attention respectively, descripted in Section 2.5. The best F1 score is written in bold.

Training was carried out for 100 epochs with a batch size of 256. The initial learning rate was tuned in [1e-3, 1e-2] and decreased according to the cosine decay rule. Adam was used as the optimizer. In the inference step, the final inference was derived by averaging the probabilities for each inferred 30 s segment. The evaluation was based on the challenge metric.

## 6 Results

### 6.1 Single Model

The performance of the top 10 models based on the macro average F1 score selected based on our validation set is reported in Table 3. All models except for M10 showed a performance of over 81%, with a difference of up to 0.6%. The best model showed a performance of 84.56% and was best for AF, I-AVB, LBBB, and STE. The performance by class exhibited a similar trend in all models. RBBB was the class best classified, with a performance of 94% or more in all models, and performance on RBBB, LBBB, and AF was more than 92% in almost all models. The PVC, STD, I-AVB, and SNR were more than 80%, PAC was more than 60%, and STE was less than 60%. Because the diagnosis of STE is influenced by the doctor's experience and subjectivity [46], this may introduce a label noise. This is supported by the fact that STE had the largest performance variance among the models.

The combinations of the components used in each model are listed in Table 3. The ResNet-18 backbone and BiGRU with lead-axis unfolding were selected for all models. This suggests that the model complexity of ResNet-34 is too high for an ECG. It is presumed that the GRU is always adopted owing to its robust nature in noisy ECG [32]. Lead-axis unfolding is adopted because it is bidirectional and considers the relationship between leads well. In the case of time-axis unfolding, the performance improvement seems to have failed because of the excessive concentration of the local relations. Referring to Chen et al., who stated the different roles of each lead, the lead axis seems to be successful because inter-lead characteristics are more important than intra-lead characteristics for the analysis of arrhythmias. Attention was used in all the top 10 models, and element attention (6 out of 10) prevailed over instance attention (4 out of 10). The SE block performed better when not used (4 four of the top five). This may be because small-scale features were lost as the reduction ratio increased. It can be observed that r=2 is adopted in all cases, even when the SE block is used.

### 6.2 Ensemble Model

Based on the top 10 models, we compared all ensemble combinations of 1013 cases, ranging the number of models from two to ten. The performance of our test set for the top five ensembles is presented in Table 4. The performance on the hidden test set of the ensemble is recorded in Table 5, which further contains the hidden test set results (CPSC2018 Top5) reported in CPSC2018. F_XX represents the average performance of a group. F_AF contains AF; F_Block contains I-AVB, LBBB, and RBBB; F_PC contains PAC and PVC; and F_ST contains STD and STE.

As a result of the experiment on our test set (Table 4), the ensemble formed based on five to seven models exhibited the best performance (avg. F1 84.35 ~ 86.47%). This suggests that the optimal number of models was approximately six. The result that using 10 models did not show optimal performance is consistent with the law of diminishing returns [47]. The performances of E01 and E05 differed by 2.12%, but E05 (84.35%) exceeded all single models except for one (M01; 84.56%). Therefore, the effectiveness of classifier fusion based on multiple views was confirmed.

For the hidden test set, the five ensembles showed results lower by 0.5~3.6% than our test set (Table 5). However, the difference between ensembles was reduced (2.12% to 1.1%), and all ensemble performances exhibited second- to third-place performances. This suggests that the multi-view performs a stable inference. The best ensemble was E05, which achieved a state-of-the-art performance of 84%. E05 had the lowest average F1 in our test set, but E05 generalized well because the performance in our test set and the hidden test set was almost the same (0.5%).

| case | Model IDs | F1 |
|---|---|---|
| E01 | 2,3,4,5,6,9 | **0.8647** |
| E02 | 2,3,4,5,7,8,9 | 0.8598 |
| E03 | 2,3,4,8,9 | 0.8574 |
| E04 | 1,2,3,4,9 | 0.8557 |
| E05 | 1,2,4,5,10 | 0.8435 |

**Table 4. Top5 ensemble performances with our test set.** Model IDs indicate the model numbers shown in Table 3. The best F1 score is written in bold.

| | Case | F_AF | F_Block | F_PC | F_ST | F1 |
|---|---|---|---|---|---|---|
| Our Models | E01 | 0.922 | 0.913 | 0.819 | 0.772 | 0.829 |
| | E02 | 0.927 | 0.912 | 0.818 | 0.781 | 0.833 |
| | E03 | 0.922 | 0.909 | 0.816 | 0.778 | 0.830 |
| | E04 | 0.920 | 0.912 | 0.817 | 0.782 | 0.831 |
| | E05 | 0.925 | **0.915** | 0.828 | **0.790** | **0.840** |
| CPSC2018 Top Models[2] | Chen et al. | **0.933** | 0.899 | **0.847** | 0.779 | 0.837 |
| | Cai et al. | 0.931 | 0.912 | 0.817 | 0.761 | 0.830 |
| | He et al. | 0.914 | 0.879 | 0.801 | 0.742 | 0.806 |
| | Yu et al. | 0.918 | 0.890 | 0.789 | 0.718 | 0.802 |
| | Yan et al. | 0.924 | 0.882 | 0.779 | 0.709 | 0.791 |

**Table 5. Top5 ensemble performances with the hidden test set.** Model IDs indicate the model numbers shown in Table 3. "CPSC2018 Top Models" refers to the performances of previous works posted on the official webpage. Since their paper submission was not mandatory, some of them have no papers published. The best F1 score in each column is written in bold. F_AF, F_Block, F_PC, and F_ST indicate each F1 score for AF, heart blocks (I-AVB, LBBB, and RBBB), premature contractions (PAC and PVC), and ST-segment abnormality (STD and STE), respectively. Grouping classes into 4 types, except for "normal", follows the official evaluation of the CPSC2018.

For our test set, we analyzed the trade-off according to the number of models in an ensemble (Figure 3). Intuitively, it was confirmed that the number of models and the performance were proportional. Because the interquartile range (IQR) decreases as the number of models increases (from 1.32 to 5.65e-1%), it can be inferred that the greater the number of models, the more stable it is. The optimal number of models for the performance was three. For #models=3, the ensemble showed the largest performance increase (2.8e-1%) and the maximum g-mean (0.7046).

---
[2] http://2018.icbeb.org/Challenge.html

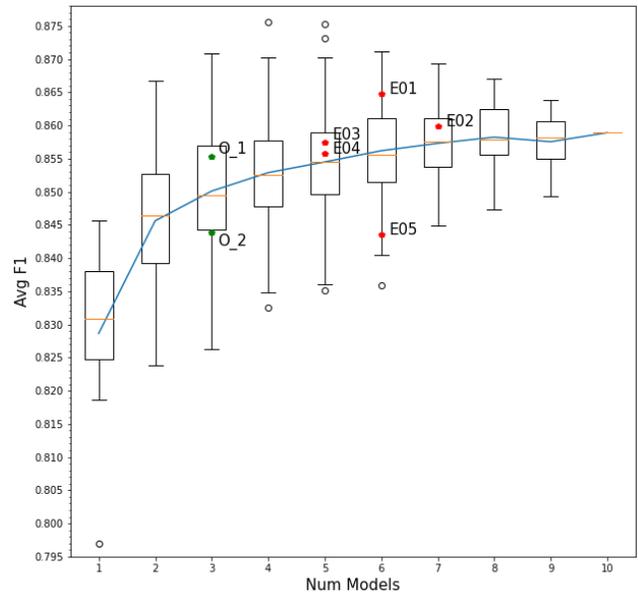

**Figure 3. F1 score on our test set with respect to the number of models in ensembles.** Ensemble performance for E01~E05 in Table 4 is marked as red dots. O_1 and O_2 indicate ensembles of #models=3, presumed mutually independent.

The results for our test sets E01–E05 are shown as red dots in Figure 3. All four ensembles, except E05, outperformed the average of the ensembles with the same number of models. However, considering that the best in the hidden test set was E05, E01–E04 could be fitted only to the public data.

However, according to Bonab et al. [47], an optimal ensemble requires independence between models. This implies that M02 and M04 adopted in all ensembles are independent and complementary. M09 was used for E01–E04, whereas M10 was used for E05 instead of M09. This indicates that (M02, M04, M09) is the optimal model set for our test set and (M02, M04, M10) is that for the hidden test set. In Figure 3, (M02, M04, M09) is denoted as O1, and (M02, M04, M10) is denoted as O2. The relationship between O1 and O2 is similar to that between E01, E04, and E05. This indicates that the independence and complementarity between the models were consistent throughout the experiment.

## 7  Conclusion

The introduction of DL significantly improved the automatic ECG diagnosis performance. However, previous studies tended not to focus on the difference between the combination of various DL components and tended to separate the artifacts from the model that occur frequently in ECGs. Therefore, we constructed and compared ensembles of models, each of which analyzes ECGs from different perspectives based on multi-view learning. In addition, we suggested augmentation specialized for ECG, leading the model to observe more diverse data. Experiments demonstrated that the integration of models can outperform state-of-the-art performance. Overall, the results of this study demonstrate the possibility of further improvements on our approach by incorporating a richer combination of DL components.